\documentclass[12pt]{article}

\usepackage{amsmath}
\usepackage{amssymb}
\usepackage{amsmath}
\usepackage{color}
\usepackage{graphicx}
\usepackage{epstopdf}
\usepackage{subcaption}
\usepackage{psfrag}
\usepackage{float}
\usepackage{hyperref}

\renewcommand{\div}{\mathrm{div}}

\newcommand{\df}{\partial}
\newcommand{\rd}{\mathrm{d}}

\def\bfq{{\bf q}}

\allowdisplaybreaks
\begin{document}

\title{Velocity of viscous fingers in miscible displacement}
\author{F.\,Bakharev,
A.\,Enin,
A.\,Groman,
A.\,Kalyuzhnuk,
S.\,Matveenko,\\
Yu.\,Petrova,
I.\,Starkov,S
S.\,Tikhomirov$^*$}


\maketitle

\begin{abstract}
The paper investigates the linear growth of the mixing zone during polymer slug injection into a water reservoir. The velocities of the slug front and of the boundaries of the mixing zone are analyzed as key parameters. Using two different numerical methods (finite volumes and finite elements), the impact of the slug size, reservoir dimensions, Peclet number, and viscosity curve shape on the corresponding velocities is examined. Notwithstanding the realization of the solution by two computational schemes, the simulation results coincide with sufficient accuracy. The numerically obtained velocities are compared with theoretical estimates within the transverse flow equilibrium approximation and Koval model. Based on the comparison pattern, recommendations are presented on the use of specific analytical methods for estimating the growth rate of the mixing zone depending on the characteristics of the polymer.
\end{abstract}

\textbf{Keywords:}
viscous fingers, porous media, water/polymer interface, miscible displacement, horizontal wells,  mixing zone, Koval model, transverse flow equilibrium, COMSOL, MRST


\section{Introduction}\label{intro}
\begin{figure}[t!]
    \centering
    \begin{subfigure}{0.39\textwidth}
        \includegraphics[width=\textwidth]{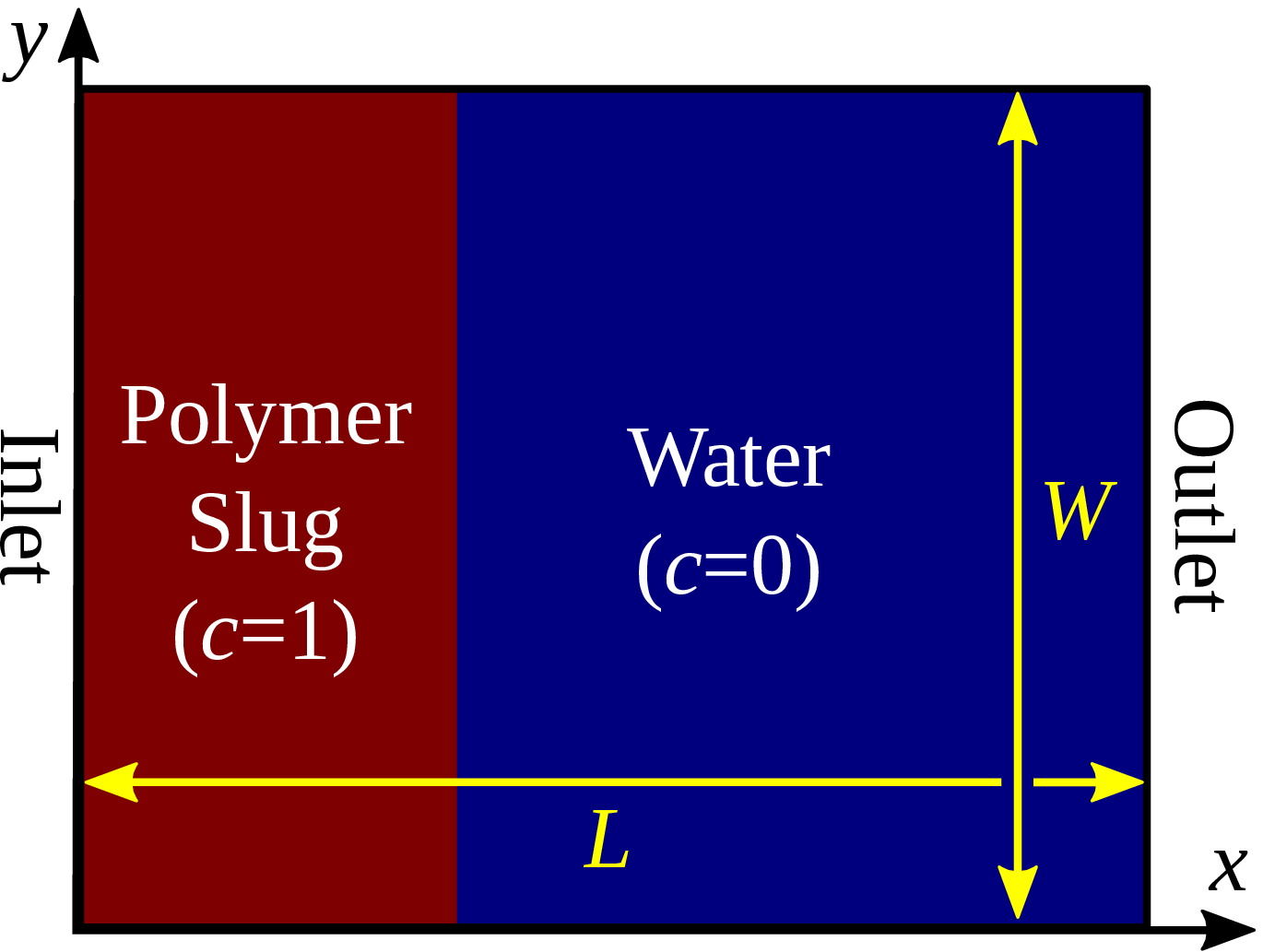}
        \caption{}
    \end{subfigure}\hspace{5mm}
    \begin{subfigure}{0.39\textwidth}
        \includegraphics[width=\textwidth]{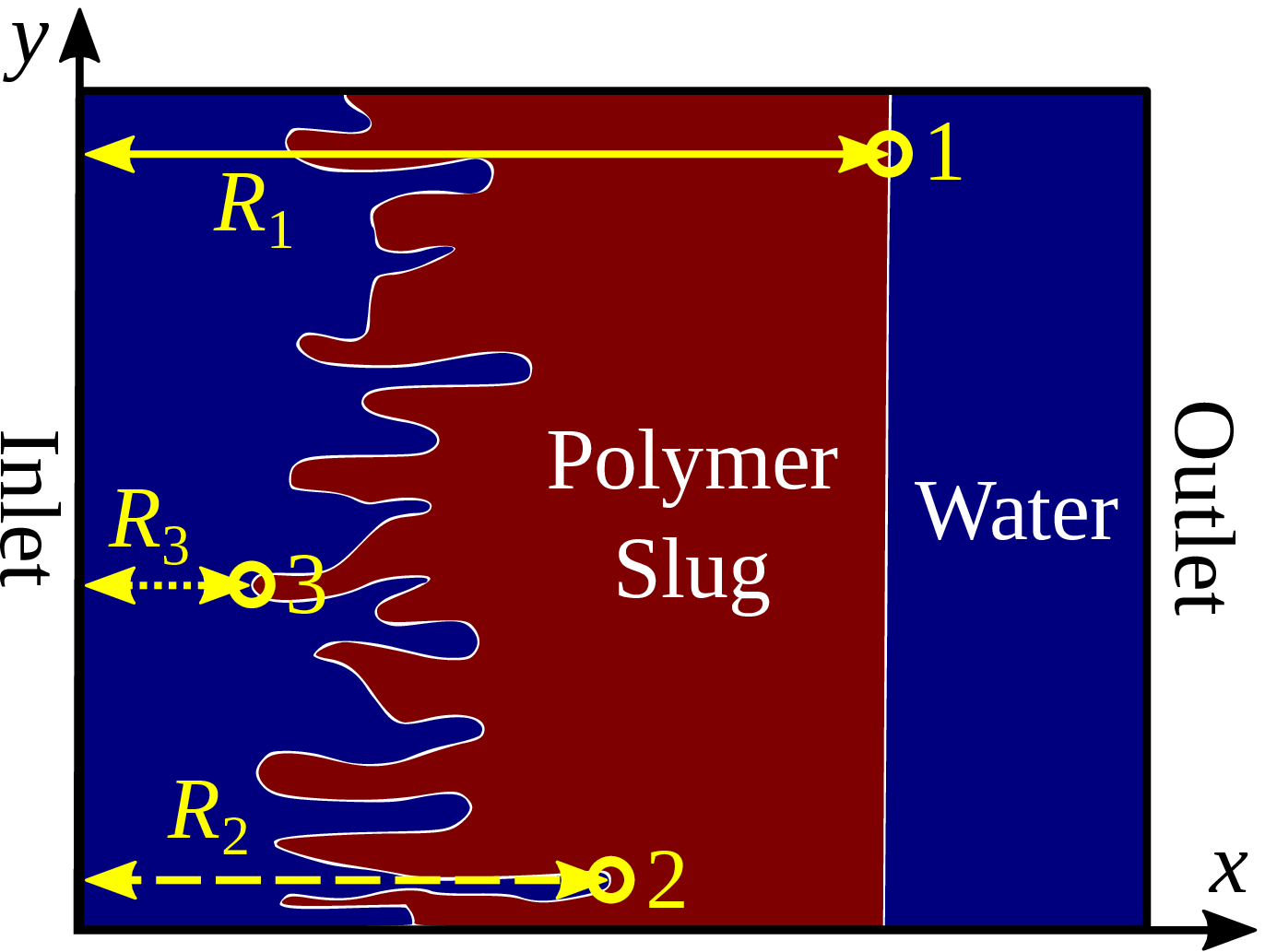}
        \caption{}
    \end{subfigure}
    \caption{The schematic representation of a polymer flooding reservoir at the initial stage (a) and  after some injection time (b). Points 1,2,3 correspond to front/rear ends of the polymer slug and the tip of the longest water finger, respectively. Thus, the distance $R_2-R_3$ defines the width of the mixing zone. The parameters $L=40$\,m and $W=31.4$\,m were taken as dimensions of computational domain.}
    \label{fig:slug-model}
\end{figure}

The study of miscible and immiscible processes in porous media attracts the researchers attention for enhanced oil recovery technologies, including water, gas or chemical flooding (polymer/surfactant) and other techniques. An extensive literature review both on various aspects of immiscible and miscible displacements can be found for instance in \cite{bakharev2020} and references therein, see also \cite{reviewmiscible2017,homsy1987,chen1998part1,chen1998part2}. The flow in the interface area during an immiscible and miscible displacement process is of special interest in the systems where interfacial properties are critically concerned, e.g.\,the production of crude oil from naturally occurring reservoirs. Small perturbations at the interface can grow into nonlinear structures having the form of fingers.
These viscous fingers can be classified as the interface instabilities in both the porous medium and the fluid flow. They are usually obtained when the less viscous fluid displaces the more viscous fluid, e.g.\,water and polymer. The way the fingers grow is a result of the combined effect of many factors, such as the mobility ratio, diffusion, gravity and the heterogeneity of the porous medium. Due to the difference in viscosity of water and polymer, the movement of the viscous front becomes unstable and forms a mixing zone Fig.\,\ref{fig:slug-model}. Such an area has a characteristic width defined by the the tip of the longest water finger and the tail end of the polymer. A detailed study of this process is one of the main aims of this work. In order to address the most realistic situation from a practical point of view, we model the process of polymer slug injection. 

An exact theoretical prediction for a fingering pattern in any given displacement is not feasible because of the extremely complicated nature of the processes involved. On the other hand, numerical modelling of viscous fingers pose a big challenge for reservoir simulation as the computational grid size must be comparable to the finger wavelength \cite{ChuokeMeursPoel5, luo2018scaling, luo2017interactions}. This demand results in excessive computational costs. As a possible remedy, engineers employ different empirical models (e.g.\,\cite{Koval15, otto2005,yortsos2006,ToddLongstaff16, Fayers17, TardyPearson18}) calibrated using rigorous simulations. Nonetheless, a detailed and comprehensive comparison between these models and simulations have not yet been reported in literature.

In our research, we are interested in the time evolution of the mixing zone size for miscible displacement. Because of this, we have identified two general approaches providing the upper estimates for the growth of the mixing zone under different assumptions. These are the transverse flow equilibrium (TFE) approximation and the Koval model with its variations. A series of simulations to compare both Koval and TFE predictions with the numerical results was carried out. Two different solvers (finite volume, finite element) were used to deal with the reliability of the proposed methodology and implementation.

The rest of the paper is organized in four parts.  The next Section\,\ref{mathematics} contains the mathematical formulation used to describe transport processes in porous media. Treating of two empirical models in detail is of particular interest in this section. In Section\,\ref{numerics} we introduce the numerical framework adopted to investigate the viscous fingering phenomenon. Specifically, the main features of COMSOL and MRST solvers are presented. Section\,\ref{mainresults} presents the results of the performed  numerical simulations. The last section, Section\,\ref{sec:summary}, has our final remarks and a description of some open problems.

\section{Mathematical formulation}\label{mathematics}
\subsection{System of basic equations}

The equations governing the time-dependent incompressible flow generated by a miscible displacement process are the Darcy law for porous media, mass balance and continuity in pressure/fluxes conditions (also known as the Peaceman model)

\begin{equation}
\begin{aligned}
  &\text{conservation of species}&&
\phi\cfrac{\df  c}{\df t}+ \div (\bfq\cdot c)=D\Delta c ,
\\
&\text{incompressibility condition}
&& \div(\bfq)=0,
\\
  &\text{Darcy's law}
  &&\bfq=-k\cdot m(c)\nabla p.
\end{aligned}\label{eq:A}
\end{equation}
Here $c$,$p$,$\bfq$ are the polymer concentration,  pressure, and linear velocity, respectively.
The mobility of the water phase $m(c)$ is inversely proportional to the viscosity $\mu(c)$, which is a known and monotonically increasing function of concentration $c$. As our attention focuses on the two-dimensional flow in horizontal plane, gravity does not have an effect. In our investigations we consider only a constant permeability and molecular diffusion. 
The latter one can be expressed as a scalar diffusion coefficient $D$. To model unstable flow behavior, we examine the system \eqref{eq:A} in a linear geometry $(x,y)\in [0,L]\times [0,W]$ with the following boundary conditions (see, Fig.\,\ref{fig:slug-model})

\begin{center}
\begin{tabular}{cccc}
    inlet, $x=0$& $c=0$ &$\bfq = q_0 \bf{e}_1$ \vspace{0.2cm}\\
     outlet, $x=L$ & $ \cfrac{\partial c}{\partial x}=0$& $p=0$\vspace{0.2cm}\\
     no-flow condition, $y=\{0,W\}$& $ \cfrac{\partial c}{\partial x}=0$& $\bfq\cdot {\bf e_2}=0$
\end{tabular}
\end{center}
By performing standard dimensionalization

\begin{equation}
    \tilde{x}=\cfrac{x}{L},\qquad  \tilde{y} = \cfrac{y}{L}, \qquad \tilde{t}=\cfrac{q_0}{\phi L} t,\qquad \tilde{\bfq}=\cfrac{\bfq}{q_0}, \qquad \tilde{p}=\cfrac{k}{q_0 L}p,
\end{equation}
and introducing the Peclet number $\mathrm{Pe}=Lq_0/D$, we obtain a new system of equations

\begin{equation}
    \begin{aligned}
     \cfrac{\df  c }{\df t}+ \div (\bfq\cdot c)=\cfrac{1}{\mathrm{Pe}}\Delta c ,\\
     \div(\bfq)=0,\\
     \bfq=-m(c)\nabla p,
\end{aligned}\label{eq:B}
\end{equation}
with unit velocity at inlet $\bfq = \bf{e}_1$ and $(x,y)\in[0,1]\times[0,W/L]$. In the equations above and in the rest of this paper we omit tildes and always use dimensionless quantities if not otherwise stated.
A number of approaches to handle the system \eqref{eq:B} analytically were proposed for the scenario with the initial condition $c=1$, i.e.\,the whole reservoir is full of polymer.

\subsection{Transverse flow equilibrium approximation}\label{TFE}

A possible simplification of the system \eqref{eq:B} under the assumption of $p(x,y,t)=p(x,t)$ and $\mathrm{Pe}\to+\infty$ was repeatedly reported in\,\cite{wooding1969,otto2005,otto2006,yortsos2006}.
In our study the latter condition reflects a small molecular diffusion. By means of the Green theorem we can write

\begin{multline}
    0=\int_{0}^{x_0}\int_0^{W/L}\div(\mathbf{q}(x,y,t)) \,dy\,dx=\\-\int_0^{W/L} q_1(0,y,t)\,\rd y+\int_0^{W/L} q_1(x_0,y,t)\,\rd y 
    =-W/L-\cfrac{\partial p}{\partial x}\int_0^{W/L} m(c(x,y,t))\,\rd t,
\end{multline}
which gives us

\begin{equation}
\begin{aligned}
    \cfrac{\partial p}{\partial x}(x,t)=-\cfrac{1}{\overline{m}(x,t)}, \qquad \text{where}\qquad \overline{m}(x,t)=L/W\cdot\int\limits_0^{W/L} m(c(x,y,t))\,\rd y.
    \label{eq:G}
\end{aligned}
\end{equation}

\noindent
Substitution of (\ref{eq:G}) into \eqref{eq:B} yields

\begin{equation}
\begin{aligned}
   &\text{conservation of species}&&
\cfrac{\df  c }{\df t}+ \div (\bfq\cdot c)=\cfrac{1}{\mathrm{Pe}}\Delta c ,
\\
&\text{incompressibility condition}
&&
\div(\bfq)=0,
\\
 &\text{velocity according to TFE}
 &&\bfq=(q_1,q_2),\quad q_1=\cfrac{m(c)}{\overline{m}(x,t)}.
\end{aligned}\label{eq:C}
\end{equation}

\noindent
Because of the absence of the pressure term, the derived system \eqref{eq:C} is easier for investigation.
A similar method for estimating the width of the mixing zone was developed for gravity driven miscible fingering in\,\cite{otto2006}. 
The no-gravity case for linear geometry was presented almost at the same time in\,\cite{yortsos2006}. 
As $m(c)$ is a monotonically decreasing function, the main idea behind these works is to estimate the velocity of the coordinate $x$ that satisfies the inequality, as follows

\begin{equation}
    \cfrac{m(c)}{m(0)}\leqslant \cfrac{m(c)}{\overline{m}(x,t)}\leqslant \cfrac{m(c)}{m(1)}.
\end{equation}

By using this concept, it is possible to construct one-dimensional barrier solutions (a kind of generalized solutions which are natural for the flows we analyze) for the system \eqref{eq:C}. For this purpose, let us consider a solution $c^{*}=c^*(x,t)$ of the one-dimensional problem

\begin{equation}
    \cfrac{\partial c^{*}}{\partial t} + \cfrac{m(c^{*})}{m(0)}\cdot \cfrac{\partial c^{*}}{\partial x} = \cfrac{1}{\mathrm{Pe}}\cfrac{\partial^2 c^{*}}{\partial x^2},
    \label{D1}
\end{equation}

\noindent
with the initial conditions $c^{*}(x,0)\geqslant c(x,y,t)$. Then, $c^{*}(x,t)\geqslant c(x,y,t)$ is fulfilled for all times\,\cite{otto2006}.
The upper barrier solution $c^*(x,t)$ majorizes the solution of the system \eqref{eq:C}. Analogously, the lower barrier solution $c_*=c_*(x,t)$ is defined by

\begin{equation}
    \cfrac{\partial c_{*}}{\partial t} + \cfrac{m(c_{*})}{m(1)}\cdot \cfrac{\partial c_{*}}{\partial x} = \cfrac{1}{\mathrm{Pe}}\cfrac{\partial^2 c_{*}}{\partial x^2},
    \label{E1}
\end{equation}
that minorizes the solution of the system \eqref{eq:C}. That is, if $c_{*}(x,0)\leqslant c(x,y,t)$ then for all times $c_{*}(x,t)\leqslant c(x,y,t)$. 

Equations \eqref{D1} and \eqref{E1} can be rewritten as

\begin{equation}
    \cfrac{\partial c^{*}}{\partial t} + \cfrac{\partial F^*(c^{*})}{\partial x} = \cfrac{1}{\mathrm{Pe}}\cfrac{\partial^2 c^{*}}{\partial x^2},
    \quad\quad
    \cfrac{\partial c_{*}}{\partial t} + \cfrac{\partial F_*(c_{*})}{\partial x} = \cfrac{1}{\mathrm{Pe}}\cfrac{\partial^2 c_{*}}{\partial x^2},
    \label{F1}
\end{equation}
for 

\begin{equation}
    F^*(c^*)=\cfrac{1}{m(0)}\int\limits_{0}^{c^*} m(c)\,\rd c,
    \quad\quad
    F_*(c_*)=\cfrac{1}{m(1)}\int\limits_{0}^{c_*} m(c)\,\rd c.
\end{equation}
The functions $F^{*}$ and $F_{*}$ are concave as $m(c)$ is a monotonically decreasing with respect to $c$. From it follows that solutions of \eqref{F1} are travelling waves with velocity given by the Rankine-Hugoniot conditions (as $\mathrm{Pe} \to \infty$)

    \begin{equation}
\begin{aligned}
    & v^\mathrm{TFE}_b:=\cfrac{1}{m(0)}\int\limits_{0}^{1} m(c)\,dc=\mu(0)\int\limits_0^1\cfrac{\rd c}{\mu(c)},
    \\
    & v^\mathrm{TFE}_f:=\cfrac{1}{m(1)}\int\limits_{0}^{1} m(c)\,dc=\mu(1)\int\limits_0^1\cfrac{\rd c}{\mu(c)}.
\end{aligned}\label{TFE-estimates}
\end{equation}
These expressions provide estimates for the velocities of the front $\left(v^\mathrm{TFE}_b\right)$ and tail $\left(v^\mathrm{TFE}_f\right)$ ends of the mixing zone and represent the TFE model.

\subsection{Koval model}\label{Koval}

Apart from mathematically rigorous methods, a number of empirical models of various sophistication have been proposed to predict the movement of the mixing zone, differing in the number of fitting parameters needed for their application. The simplest and widespread
approach for evaluating the mean transport of the solvent through the reservoir is the Koval model\,\cite{koval1963} and its variations -- Booth (naive Koval)\,\cite{booth2008phd,booth2010} and Todd-Longstaff \cite{toddlongstaff1972} models. A common feature of these methods is the use of a key parameter, an effective mobility ratio. In all the cases the dynamical equation is

\begin{equation}
    \cfrac{\partial\bar{c}}{\partial t}+ \cfrac{\partial}{\partial x}\left(\cfrac{\bar{c}}{M_e-(M_e-1)\bar{c}}\right)=0.
    \label{eq:Koval}
\end{equation}
Here, $\bar{c}$ is the solvent concentration averaged in the direction perpendicular to the flow. An effective mobility ratio $M_e$ can be defined as

\begin{center}
\begin{tabular}{cc}
    Booth (naive Koval model)  & $M_e=M:=\cfrac{\mu(1)}{\mu(0)}$,\vspace{2mm}\\
     Koval & $M_e=(a\cdot M^{1/4}+(1-a))^4$,\vspace{2mm}\\
     Todd-Longstaff & $M_e=M^{\omega}$.
\end{tabular}
\end{center}
The parameter $a$ is typically equal to 0.22 in the Koval model\,\cite{koval1963}. The miscibility parameter $\omega$ is a fitting parameter in the Todd-Longstaff approach. Because of it, we omit this model from the remaining discussion.

\subsection{Comparison of TFE and naive Koval estimates}

Following the naive Koval model, we can write the estimates on the velocity of front $\left(v^{K}_b\right)$ and back $\left(v^{K}_f\right)$ end of the mixing zone as

\begin{equation}
     v^{K}_b = \cfrac{1}{M} = \cfrac{\mu(0)}{\mu(1)},\qquad
    v^{K}_f = M = \cfrac{\mu(1)}{\mu(0)}. 
\end{equation}

\noindent
Then, comparison with the TFE model leads to

\begin{equation}
\begin{aligned}
    v^\mathrm{TFE}_f = \mu(1) \int_0^1 \cfrac{\rd c}{\mu(c)} \leq \cfrac{\mu(1)}{\mu(0)}  = M = v^K_f, \\
    v^\mathrm{TFE}_b = \mu(0) \int_0^1 \cfrac{\rd c}{\mu(c)} \geq \cfrac{\mu(0)}{\mu(1)}  = \cfrac{1}{M} = v^K_b.
\end{aligned}\label{naive-koval-otto}
\end{equation}

\noindent
By virtue of the monotonic character of the $\mu(c)$ function, it is easy to conclude that TFE approximation provides always lower estimates than the ones given by the naive Koval model. In other words, above expressions \eqref{naive-koval-otto} demonstrate that the naive Koval estimates are too pessimistic, so we also omit its consideration.

\subsection{Statement of the problem}

The above-described theoretical concepts can be applied for analysis of the finger growth rate at the polymer slug injection. In spite of the obvious impact of the front boundary of the slug, this information would be helpful to predict the slug breakthrough behavior.
For our purpose we consider the following problem formulation.
At the initial stage, the area near the inlet already contains a uniformly distributed polymer with a concentration of $c=1$, as sketched in Fig.\,\ref{fig:slug-model},(a). The volume of this polymer corresponds to the injected slug of a given size. The rest of the reservoir is filled with water ($c=0$). After that, water pumping from the inlet begins. Such a scenario can significantly reduce the time consuming simulations. Let us emphasize that numerous verification tests confirmed the equivalence of this scenario and the complete process of injection of a polymer slug into the water reservoir (first injection of polymer into water followed by water pumping). As a particular case, we considered the injection of water into the reservoir filled with polymer, i.e.\,when pure water is completely absent at the initial moment. Formally, this corresponds to the 1\,PV slug size.

The injection process forms a mixing zone characterized by the width, which is defined by the the tip of the longest water finger and rear end of the slug (see, Fig.\,\ref{fig:slug-model},(b)). These boundaries move with different velocities. We want to reproduce this behaviour using simulations and obtain the estimates for these velocities by the Koval and TFE approaches. Moreover, it is commonly believed\,\cite{Oswald97,Gupta74} that the growth of the mixing zone is linear in time, which corresponds to the constancy of the corresponding velocities. Thus, it is necessary to understand what these rates depend on. Namely, whether they depend on the size of the slug, dimensions of the reservoir, Peclet number, type of the viscosity function. Further, going into details, the TFE approach has a physically sounder basis than Koval estimates. However, we do not know how the results obtained within the TFE approximation agree with the initial system of equations. Therefore, it is required to check to what extent the TFE estimates are being fulfilled and in which case they are accurate. Finally, we also want to see how good the Koval model overall is.

\section{Numerical framework\label{numerics}}
\subsection{Verification by two numerical methods}

There exist several private and open-source modelling packages able to deal with multiple physical and chemical processes (for an overview, see \cite{KeyesMcInnesWoodwardGroppMyraPerniceWohlmuth2013}).
The proper selection of the solver is of central importance and it may have strong impact on the development stages of the research. According to the aforementioned criteria, in the framework of the present paper, we choose two solvers, COMSOL Multiphysics\,\cite{Comsol38} and MRST\,\cite{MRST}. In this section, a brief overview of the main features of each software package is presented. Specifically, we highlight those characteristics which have been directly used or modified for the current study. 

\subsubsection{COMSOL}
COMSOL Multiphysics~\cite{Comsol38} is a finite element analysis and solver software for various engineering and physics problems. The package offers several modules for specific applications and, crucially, any specific value or equation can be introduced and/or modified by the user. 
As a part of this study, the system \eqref{eq:A} was implemented in COMSOL using the equation based modelling capability. The numerical implementation was accomplished by means of the PDE mode for time dependent analysis in the weak form. The linear solver MUMPS with implicit time-stepping method BDF was revealed to be the most stable to obtain the solution. For space discretization we selected first-order Lagrange elements and unstructured grid (Delaunay triangulation algorithm) with an average mesh element size of 0.17\,m.

\subsubsection{MRST}
The MATLAB Reservoir Simulation Toolbox (MRST)\,\cite{MRST} is a free distributed package for the MATLAB which was applied for the modelling of polymer flooding\,\cite{bao2017}. This toolbox provides
cell-centered finite volume method of discretization with two point flux approximation. In our investigation we use the linear solver with first-order fully implicit finite-difference scheme for the time derivative and the constant time step of 0.8\,h. These computational parameters and algorithm provide the appropriate balance between the accuracy and the cost of the simulations. The unstructured triangular mesh was build by a simple MATLAB code (DistMesh) realised by Persson and Strang \cite{Persson05}. The final grid consisted of approximately $10^5$ cells with mean length of 0.17\,m. As in the COMSOL simulation parameters, such a mesh size gives results consistent with those achieved by using a cell size reduced down to 0.1\,m, but requires less computational time.

In both simulators we employed irregular triangular meshes for initialization of the viscous fingering process. In contrast to irregular connectivity, structured grids are aligned in the flow direction resulting in loss of the unstable nature of the phenomenon and non-uniform concentration distribution.
Delaunay triangulation builds a highly branched network with a significant number of connections, that may be referred to as the number of pores connected to each node. Such irregularities or instabilities trigger the small perturbations leading to viscous fingering pattern formation. It is noteworthy to mention that changing the mesh element size does not dramatically affect the appearance of viscous fingers, as this phenomenon is mainly controlled by diffusion in the model parameters used.

\subsection{Modelling parameters}

\begin{figure}
    \centering
    \begin{subfigure}[b]{0.95\textwidth}
        \includegraphics[width=\textwidth]{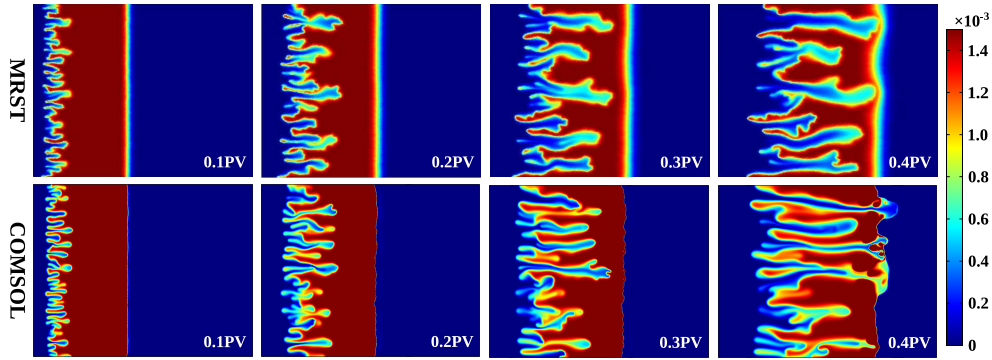}
        \caption{The polymer diffusion coefficient $D = 10^{-8}$\,m$^2$/s.}
        \label{fig:1e-8}
    \end{subfigure}
    \begin{subfigure}[b]{0.95\textwidth}
        \includegraphics[width=\textwidth]{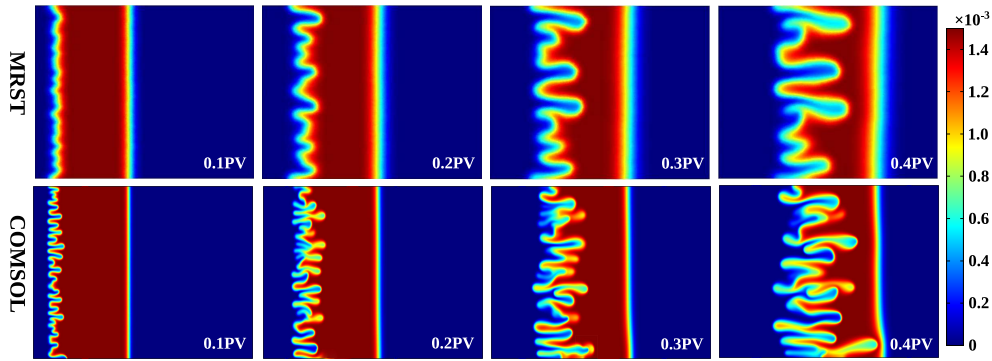}
        \caption{The polymer diffusion coefficient $D = 5\cdot10^{-7}$\.m$^2$/s.}
        \label{fig:5e-7}
    \end{subfigure}
    \caption{The comparison of the polymer concentration distribution by MRST and COMSOL packages in the case of flooding a 1/3\,PV slug at different flooding times: 0.1, 0.2, 0.3, 0.4\,PV. The results were obtained with the diffusion coefficient (a) $D = 10^{-8}$\,m$^2$/s and (b) $D = 5\cdot10^{-7}$\,m$^2$/s. The difference in slug edge tailing between MRST and COMSOL is explained in Section\,\ref{sec:pecnum}.}
    \label{fig:concentration}
\end{figure}

\begin{figure}[t!]
 \centering
 \includegraphics[width=0.49\textwidth]{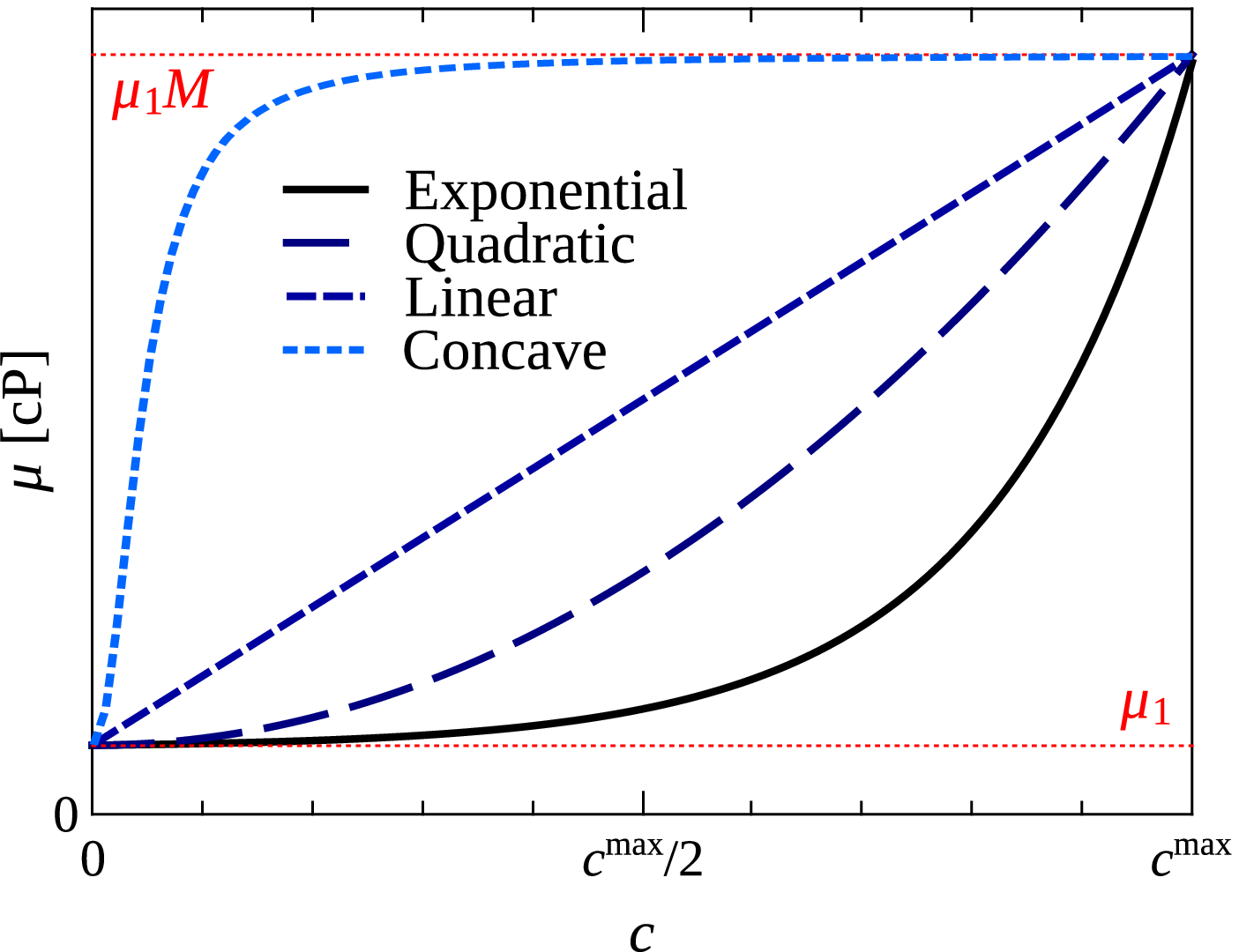}
 \caption{The different kinds of viscosity function considered in the simulations. The corresponding analytical expressions are listed in Tab.\,\ref{tab:param}. The initial polymer concentration was taken to be $c^\mathrm{max}=0.0015$ in all the simulations. The viscosity ratio parameter $M$ was varied within the range 2 to 51.}
\label{fig:vis} 
\end{figure}

The distribution of polymer concentration and its evolution over time were considered and analyzed as the output data of numerical modelling. Figure\,\ref{fig:concentration} maps a typical instance of such distribution.
Various parameters were varied in order to check the reliability of simulation results for comparison with theoretical estimates. A comprehensive series of simulations was carried out for slugs of various sizes, a set of diffusion coefficients, and a number of shapes of viscosity function (see, Fig.\,\ref{fig:vis}). Additionally, the case of water injection into the reservoir full of polymer was also studied. The parameters involved in modelling are summarised in Tab.\,\ref{tab:param}. Despite the difference in the numerical schemes employed and methods implemented into the software packages, the results of COMSOL and MRST demonstrate fairly good agreement for all considered cases, thereby confirming the general picture of physical processes based on the system \eqref{eq:A}. Let us again emphasize that all the simulations presented in this paper are for horizontal flows in which the effects of gravity can be ignored.

\begin{table}[H]
 \centering
 \begin{tabular}{c|l}
 \hline\hline
Reservoir parameters  & Data values\\
 \hline
Initial concentration of polymer, $c^\mathrm{max}$ & 0.0015\\
Permeability, $k$  & 80 [mD]\\
Viscosity of water, $\mu_1$  & 0.3 [cP]\\
Average porosity, $\phi$  & 0.19\\
Surface injection rate, $q_0$ & 10 [m$^2$/d]\\
Diffusion coefficient, $D$  & 10$^{-8}$, $5\cdot 10^{-8}$, 10$^{-7}$, $5\cdot 10^{-7}$ [m$^{2}$/s]\\
Viscosity ration parameter, $M$ & 2, 10, 20, 51\\
Computational domain length, $L$ & 40 [m]\\
Computational domain width, $W$ & 31.5 [m]\\
Linear viscosity function, $\mu$ & $\mu_1+ \mu_1 (M-1)c/c^\mathrm{max}$ [cP]\\
Quadratic viscosity function, $\mu$ & $\mu_1+ \mu_1(M-1)(c/c^\mathrm{max})^2$ [cP]\\
Exponential viscosity function, $\mu$ & $\mu_1+9.3\cdot10^{-3}(M-1)\left[\exp\left({c}/{0.00026}\right)-0.997\right]$ [cP]\\
Concave viscosity function, $\mu$ & $\mu_1+\cfrac{150(M-1)}{500+\left[(c^\mathrm{max}/c)^2)-1\right]}$ [cP]\\ \hline
 \end{tabular}
 \caption{The parameters used to model the viscous finger movement.}
 \label{tab:param}
\end{table}

\section{Results and discussion}\label{mainresults}
This portion of the paper contains the results of MRST and COMSOL simulations and its comparison with theoretical estimates. The section\,\ref{sec:fingerrate} introduces the determination procedure for the viscous fingers from the numerical solutions and the concept of what the growth rate is. Because of the significant importance of the diffusion coefficient (associated with inverse Peclet number), we analyze the influence of this parameter in section\,\ref{sec:pecnum}. The following section \,\ref{sec:slugsize} proves the independence of the finger velocity on the slug size and reservoir dimensions. The verified numerical solutions are compared to TFE and Koval estimates for different kinds of the viscosity functions in section\,\ref{sec:ottokoval}. A partial explanation of the observed patterns within the empirical models are given in section\,\ref{sec:empmodels}.

\subsection{Growth rate of viscous fingers}\label{sec:fingerrate}

There is no difficulty with a physical understanding of the concept of a viscous finger. That is, an interfacial instability phenomenon that occurs in porous media, when a high viscosity fluid is displaced by another fluid that has low viscosity. However, from the numerical side of the subject, this concept is not that clear. In order to find the growth rate of viscous fingers and velocity of the slug front by means of the simulations, it is necessary to determine the boundary position of these formations at each time step, as sketched in Fig.\,\ref{fig:slug-model}.  Formally, the tip of the fastest water finger is defined as the point within the polymer slug that is farthest from the inlet and where $c<c^\mathrm{max}$ (point 2, Fig.\,\ref{fig:slug-model}). The implementation of this inequality is often ambiguous because of the grid partition -- from the point of view of post-processing of output data. Therefore, we define the coordinate of the finger tip $R_2$ by the level of polymer concentration $c=c^\mathrm{th}=0.95c^\mathrm{max}$ to stabilize the data processing procedure. The positions of the slug boundaries can be resolved in a similar way. The front-end edge can be found by the largest distance $R_1$ to the point 1, where the polymer concentration is $c^\mathrm{th}=0.95 c^\mathrm{max}$. In turn, the rear end is found through the shortest distance $R_3$ to the point 3 with the concentration $c^\mathrm{th} = 0.05 c^\mathrm{max}$. Additional reasoning for the use of these threshold concentration levels is provided below. In the case of the water injection into polymer, only the boundaries of the mixing zone $R_{2,3}$ were examined.

\begin{figure}
    \centering
    \begin{subfigure}[b]{0.49\textwidth}
        \includegraphics[width=\textwidth]{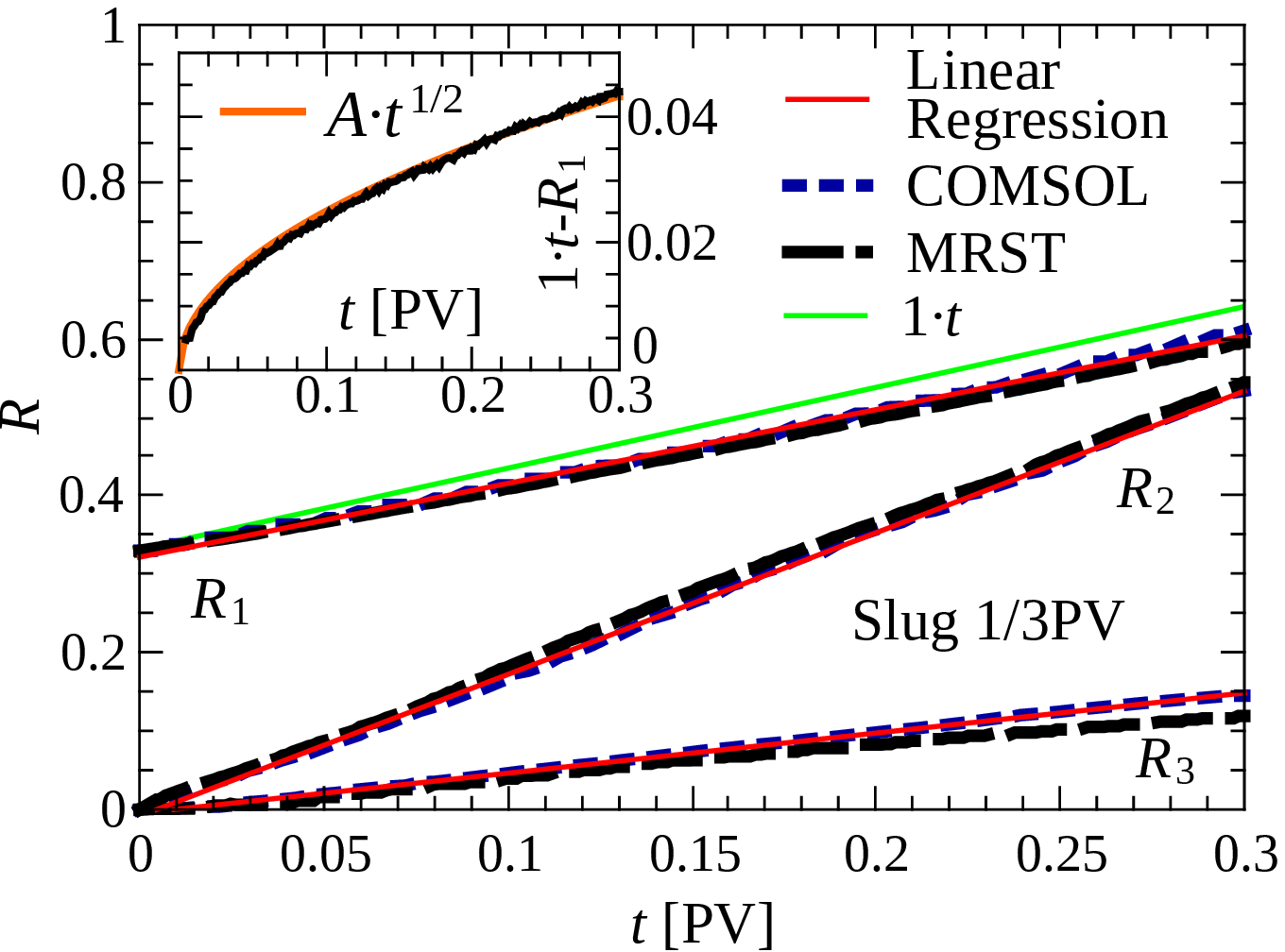}
        \caption{}
        \label{fig:R_13}
    \end{subfigure}
    \begin{subfigure}[b]{0.475\textwidth}
        \includegraphics[width=\textwidth]{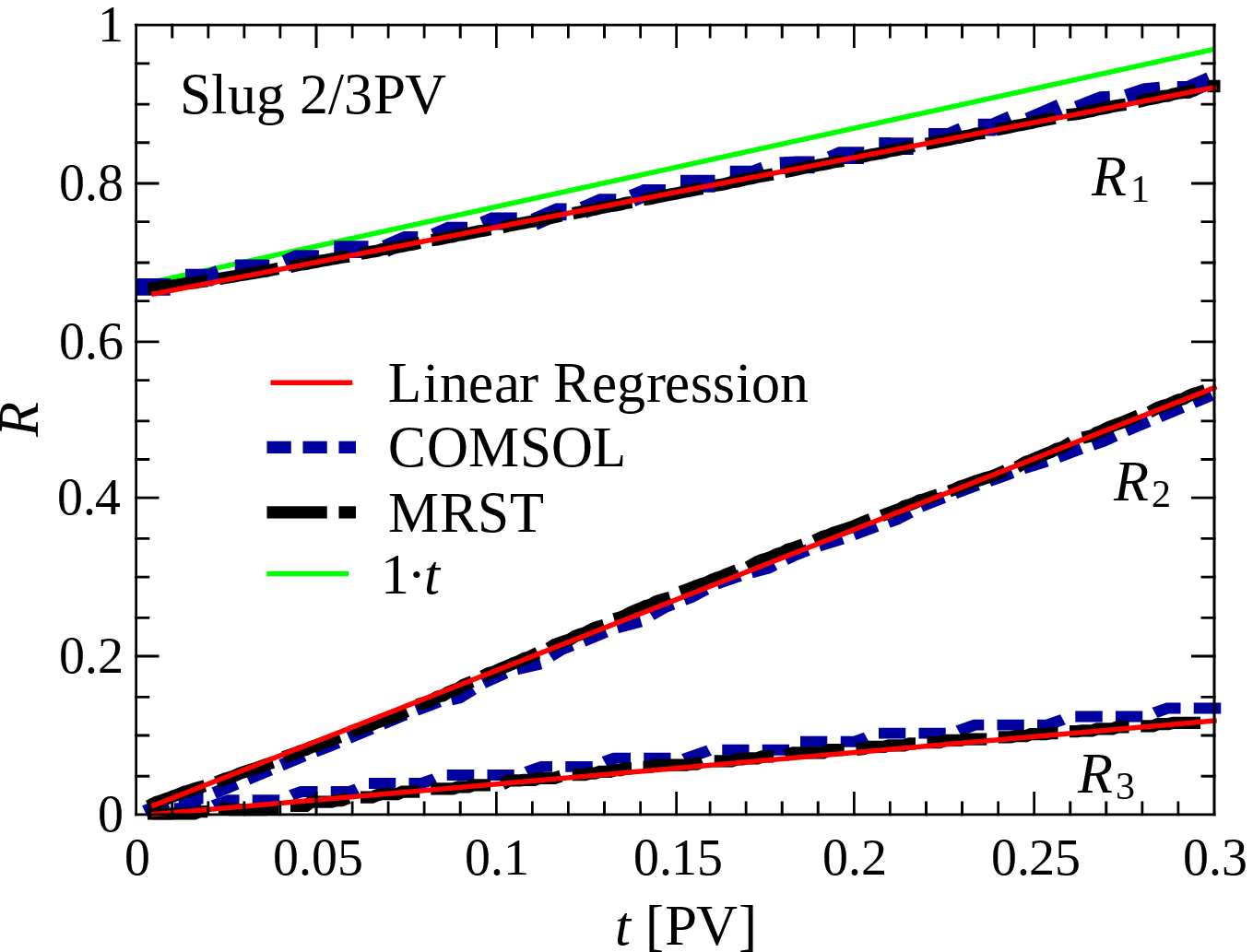}
        \caption{}
        \label{fig:R_23}
    \end{subfigure}
    \caption{The dependence of distancies between the inlet and the front $R_1$/rear $R_3$ ends of the slug and between the inlet and the outermost finger $R_2$ on the flooding time $t$. The curves were calculated with a quadratic viscosity function (see,  Fig.\,\ref{fig:vis}, $M=20$) for a 1/3\,PV (a) and  2/3\,PV (b) slug size with a diffusion coefficient $D=10^{-7}$\,m$^2$/s. Both simulators used in the calculations demonstrate a good repeatability of the results.
    Inset: the difference of the curves $1\cdot t$ and $R_1$ as a function of $t$. }
    \label{fig:R}
\end{figure}

Figure \ref{fig:R} depicts the dependence of the distances $R_{1,2,3}$ on the dimensionless time in the pore volume injected for a 1/3\,PV and 2/3\,PV slug size. For convenience, the distances were either normalized to the length of the reservoir $L$ (the distance between the inlet and outlet). All the processed curves have an almost linear character of growth, i.e.\,they reflect the constancy of the movement velocities of mixing zone and the front end of the slug. Such a linear behaviour was observed for all the performed simulations regardless of the varied parameters before the impact of the slug front on the mixing zone prevails. 
An exact analytical solution for the velocity of the slug front boundary can be adopted from a homogeneous problem of water injection. 
According to this solution, the distance $R_1$ in dimensionless notation is equal to $1\cdot t+A\sqrt{t}$, where $A$ is some coefficient depending on Peclet number. The first term of this sum is the multiplication of the unit normalized velocity (1/PV) on injection time (green solid lines in Fig.\,\ref{fig:R}). If we subtract the straight line $R = 1\cdot t$ from $R_1(t)$, then we get a root function shown in inset of Fig.\,\ref{fig:R}, 
which confirms earlier results reported in the literature. Thus, it is necessary to admit that the linear regression approximation assigns velocities to the $c^\mathrm{th}$ concentrations with some inaccuracy. It is worth to specify that the growth regime of the viscous fingers can be also different to linear at short times of injection\,\cite{Nijjer}, but investigation of such scenarios is out of the scope of our paper.

\begin{figure}[t]
    \centering
     \includegraphics[width=0.49\textwidth]{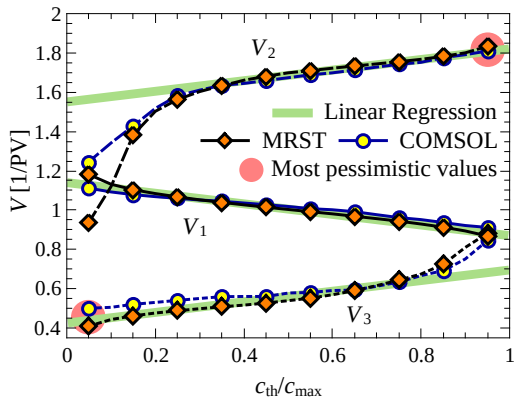}
        \caption{The movement velocity of the front $V_1$ or rear $V_3$ ends of the polymer slug and the tip of the longest water finger $V_2$ as a function of the threshold concentration level $c^\mathrm{th}$. We interpret the selected threshold concentration levels $c^\mathrm{th}=0.05,0.95$ as the most pessimistic. The results were obtained for a quadratic viscosity function (see Fig.\,\ref{fig:vis}, $M=20$) with a diffusion coefficient $D=10^{-7}$\,m$^2$/s.}
\label{fig:c_th}  
\end{figure}

The growth rate of the polymer fingers $V_{2}$ as well as the velocities of the front/rear ends of the slug are obtained by using linear regression of the data $R_{1,2,3}(t)$ via the slope coefficient. Here, it is important to note that the threshold level $c^\mathrm{th}$ affects the results when determining the distances $R_{1,2,3}$ and the subsequent finding of the velocities. Thus, Fig.\,\ref{fig:c_th} illustrates the dependence of $V_{1,2,3}$ on $c^\mathrm{th}$ parameter for COMSOL and MRST. An interesting fact is that the absolute slope of the curves (if again using linear regression) is the same for all three rates and is equal to 0.27. Such a behaviour can be traced to the presence of diffusion, both given in the system \eqref{eq:A} and numerical one.
Notwithstanding the fact that linear regression does not determine the constant velocity absolutely correctly due to the nonlinear effect of diffusion, we hold a view that the extreme concentrations only accelerate at the outermost finger and only slow down at the rear end. Accordingly, the found slopes, although they do not correspond real linear velocity, match with its upper/lower numerical estimates.

\subsection{Influence of the Peclet number}
\label{sec:pecnum}

\begin{figure}[t]
    \centering
    \begin{subfigure}[b]{0.49\textwidth}
        \includegraphics[width=\textwidth]{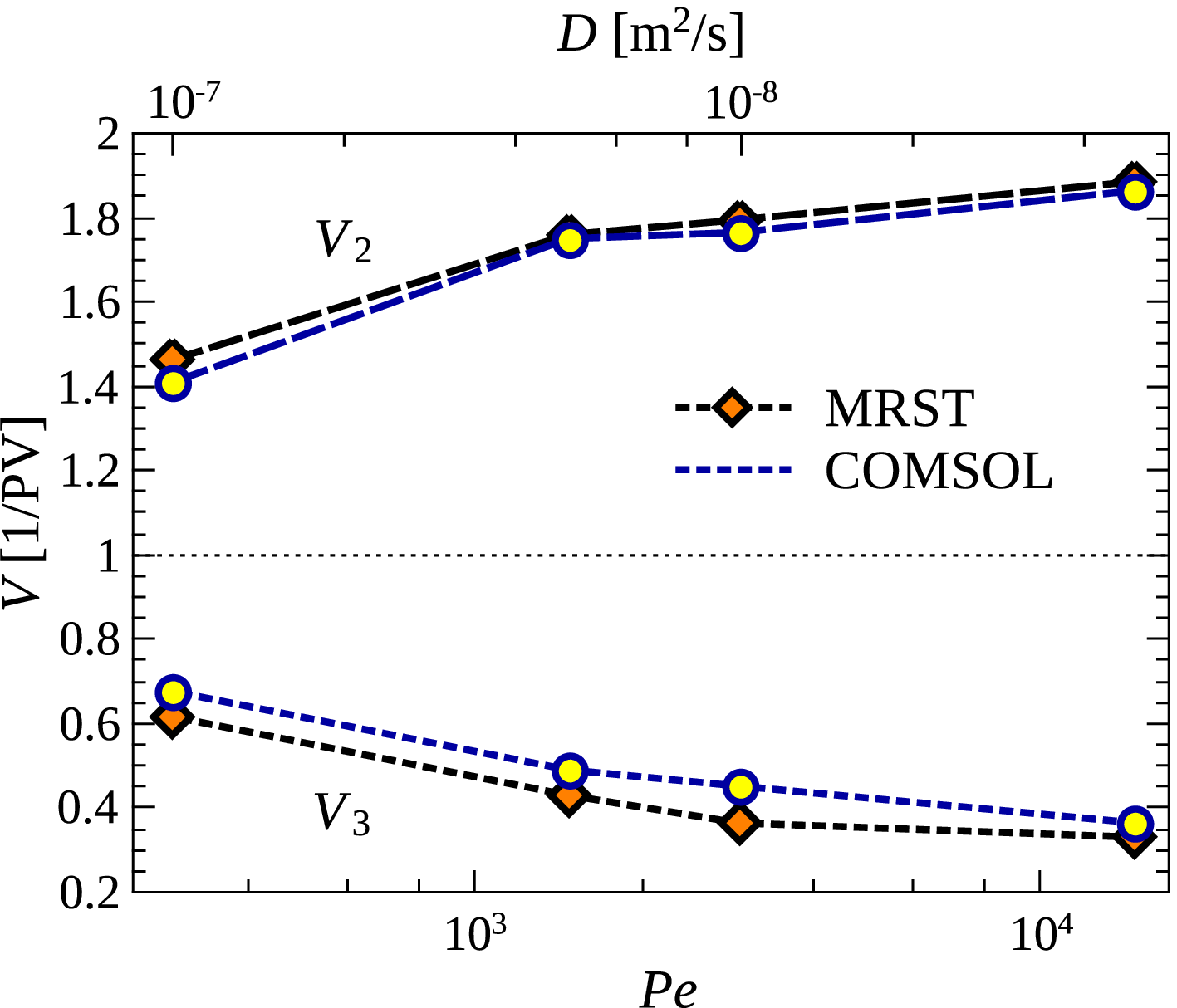}
        \caption{}
        \label{fig:V_D_w}
    \end{subfigure}\hspace{1mm}
    \begin{subfigure}[b]{0.49\textwidth}
        \includegraphics[width=\textwidth]{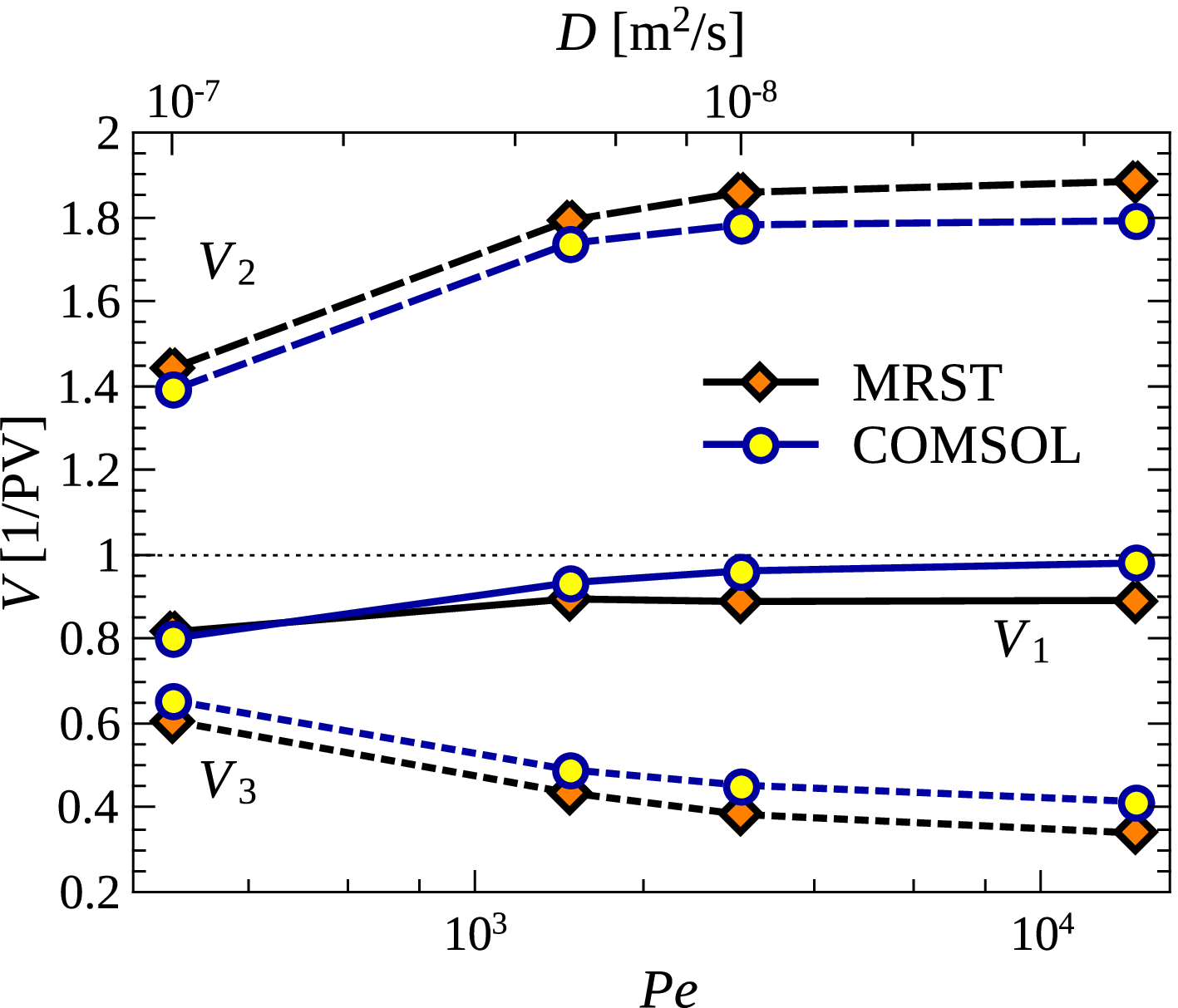}
        \caption{}
        \label{fig:V_D_slug}
    \end{subfigure}
    \caption{The velocities $V_2,V_3$ in the case of water penetration into the polymer (a) and $V_1,V_2,V_3$ for a polymer slug with a size of 1/3\,PV (b) vs.\,the Peclet number (diffusion coefficient $D$). The simulations were performed with a quadratic function of viscosity (see Fig.\,\ref{fig:vis}, $M=20$).}
    \label{fig:V_D}
\end{figure}

One of the parameters included in the theoretical estimates \eqref{eq:C} and significantly affecting the modelling results based on the system \eqref{eq:A} is the polymer diffusion coefficient $D$ or the Peclet number $\mathrm{Pe}$ following from it. To track the impact of these numbers, we ran a series of simulations and repeated the data processing described in the previous section. Simulations were run for a 1/3PV slug size and a quadratic viscosity function ($M=20$). The output results from COMSOL and MRST packages are consistent in Fig.\,\ref{fig:V_D_slug} and show clear trends.

A decrease in the Peclet number leads to a slowdown of the fingers, but increases the velocity of the slug rear-end. The effect of edge tailing is obtained as a result of the presence of dissipative terms in the system \eqref{eq:A} and numerical diffusion. Note that we may observe a difference in the width of the mixing zone obtained in COMSOL and MRST. This difference can be substantially explained by the use of stabilizing schemes for numerical diffusion in MRST, which are not available in COMSOL. When the diffusion coefficient decreases below $10^{-8}$\,m$^2$/s, COMSOL fails to obtain convergence of the solution, while the resulting velocities from MRST do not change.

We recall of the computational problem associated with the simulation of fluid dynamics by means of spatial discretization approaches. The numerical diffusion dramatically affects the results for high Peclet numbers. To ensure that numerical diffusion does not overwhelm the molecular diffusion $D$ we compared the COMSOL simulation output with the analytical solution from \,\cite{Orr07} for a  homogeneous one-phase flow. These results confirm that the employed numerical scheme always displays only small amounts of numerical diffusion across the entire range of Peclet numbers. 
 
\subsection{Independence of slug size}\label{sec:slugsize}

\begin{figure}[t]
\centering
\includegraphics[width=0.49\textwidth]{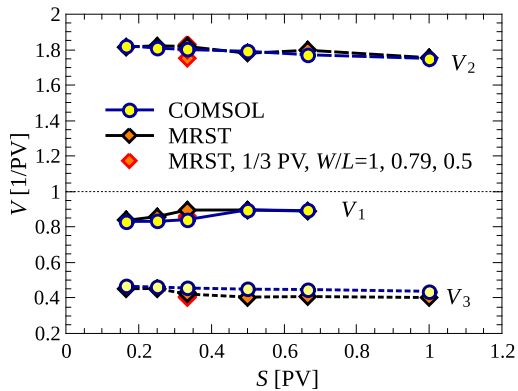}
\caption{The movement velocities of the front $V_1$\,/\,rear $V_3$ ends of the slug and the tip of the longest water finger $V_2$ for different sizes of the slug. In numerical calculations, a quadratic viscosity function was used (see Fig.\,\ref{fig:vis}, $M = 20$) and the diffusion coefficient of polymer $D = 10^{-7}$\, m$^2$/s. To eliminate the influence of the reservoir size on the simulation results, we carried out simulations in MRST for various $W/L$ ratios for a 1/3\,PV slug size (red rhombuses).}
\label{fig:V_S}  
\end{figure}

Continuing with Fig.\,\ref{fig:V_D}, we examined the results of injection for different-sized polymer slugs and pumping water into the polymer. In this setup, we were interested in the reproducibility of the results. Injection time intervals and corresponding slug sizes of 1/6, 1/4, 1/3, 1/2, 2/3\,PV were considered. Modelling was performed with a quadratic function of viscosity ($M=20$) and a polymer diffusion coefficient of $10^{-7}$\,m$^2$/s. The resulting dependencies are demonstrated in Fig.\,\ref{fig:V_S}. To ensure that there is no influence of the size of the numerical domain (reservoir area), we performed numerical calculations in MRST for various ratios $W/L=1, 0.79, 0.5$, but only for a 1/3\,PV polymer slug. These dots are also presented in Fig.\,\ref{fig:V_S} and have the same spread as other data. The difference in the velocities of the front end from 1 is associated, as we explained earlier, with a diffusion process. It can be concluded that the size of the slug has no or very little effect on the movement of the fingers and the slug boundaries. This observation is consistent with the results of the theoretical assessments.

\subsection{Comparison to TFE and Koval models}\label{sec:ottokoval}

\begin{figure}[H]
    \centering
    \begin{subfigure}[b]{0.49\textwidth}
        \includegraphics[width=\textwidth]{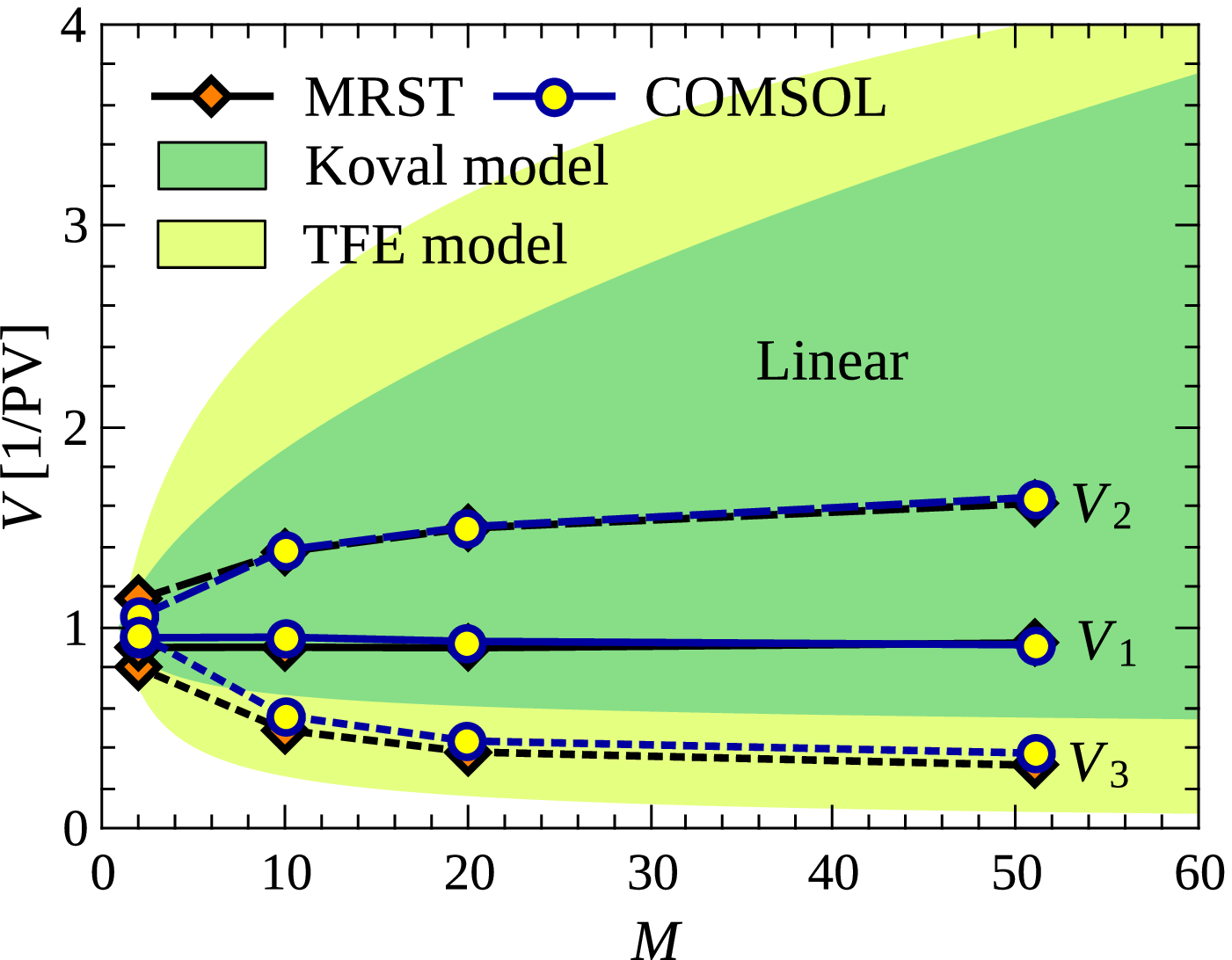}
        \caption{}
        \label{fig:V_M_lin}
    \end{subfigure}
    \begin{subfigure}[b]{0.49\textwidth}
        \includegraphics[width=\textwidth]{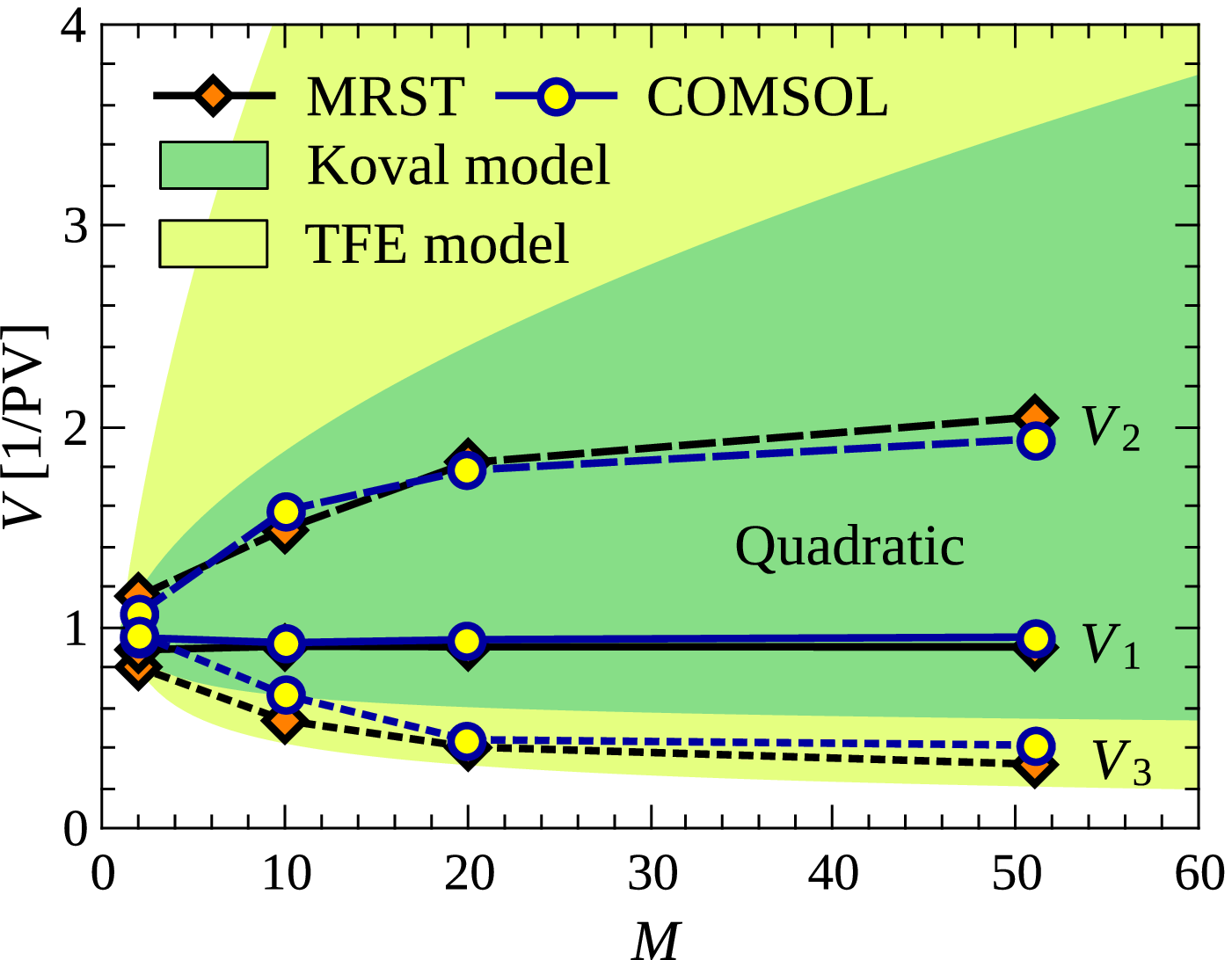}
        \caption{}
        \label{fig:V_M_quad}
    \end{subfigure}\\
    \begin{subfigure}[b]{0.49\textwidth}
        \includegraphics[width=\textwidth]{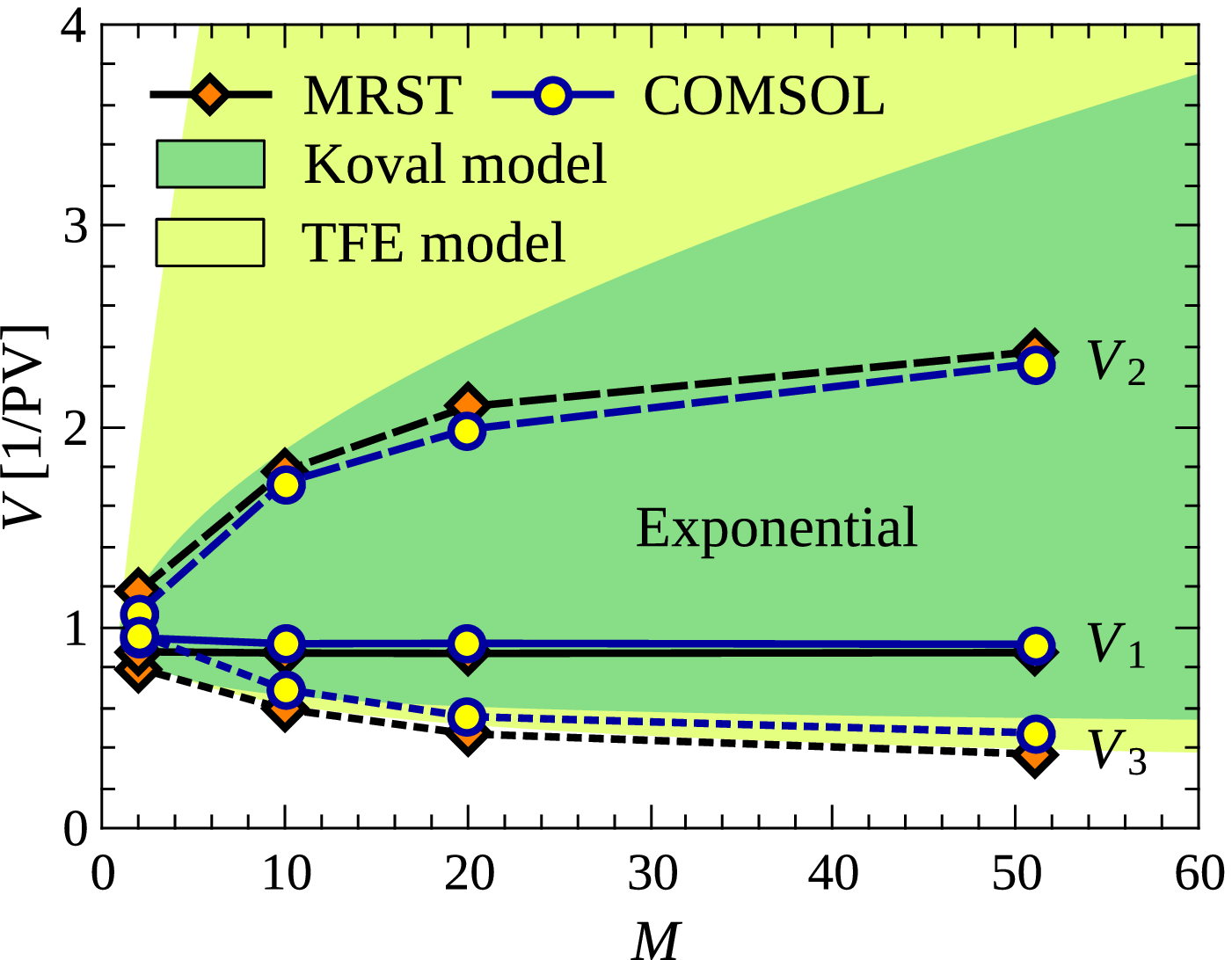}
        \caption{}
        \label{fig:V_M_exp}
    \end{subfigure}
    \begin{subfigure}[b]{0.49\textwidth}
        \includegraphics[width=\textwidth]{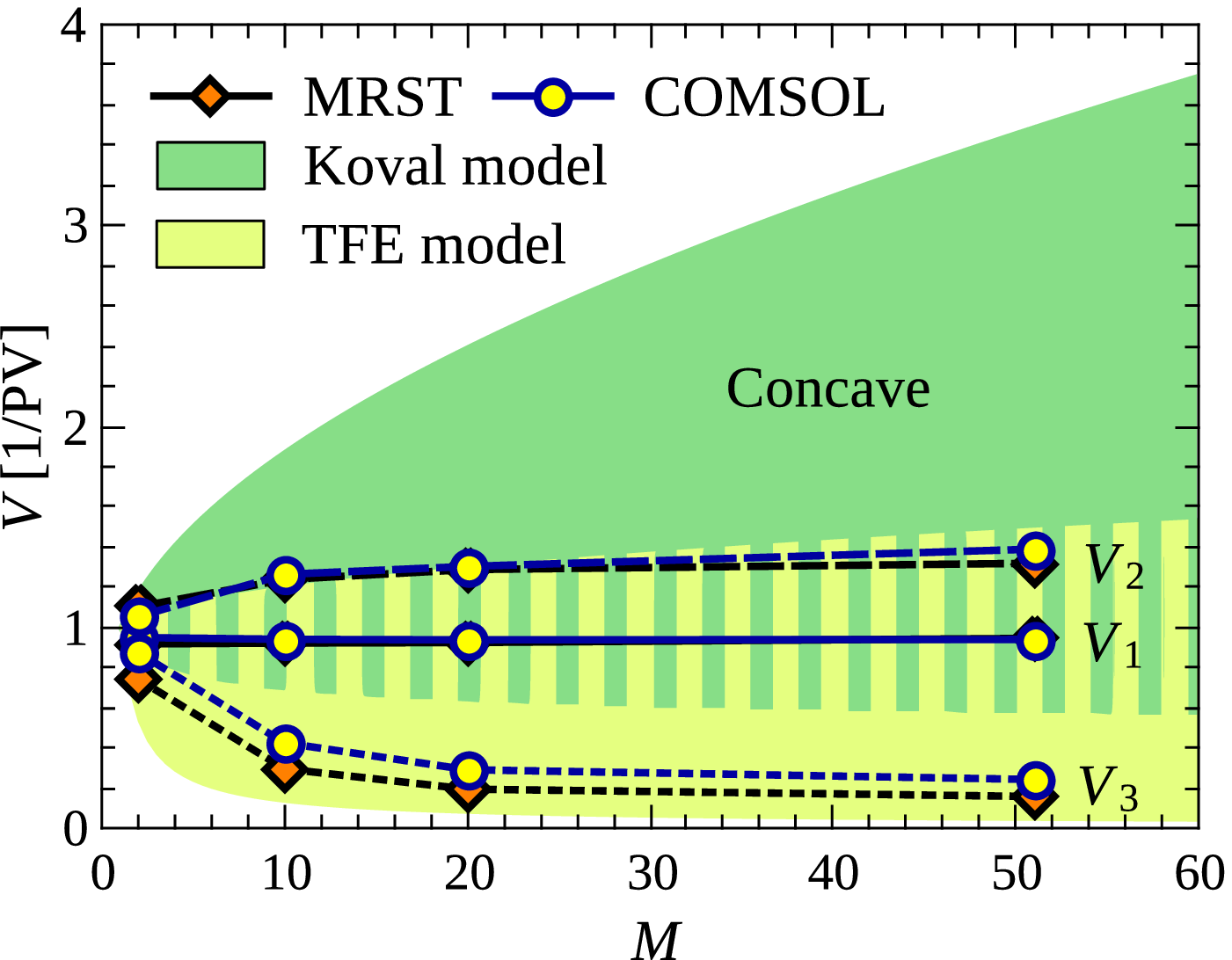}
        \caption{}
        \label{fig:V_M_con}
    \end{subfigure}
    \caption{The analysis of the influence of viscosity ratio factor $M$ on the velocities $V_1$, $V_2$ and $V_3$. For comparison, the results are given for different types of the viscosity function (see, Fig.\,\ref{fig:vis}) -- (a) linear, (b) quadratic, (c) exponential (d) concave. The diffusion coefficient of polymer $D = 10^{-7}$\, m$^2$/s was used in the simulations. In addition, the boundaries of the mixing zone defined by the TFE and Koval models are highlighted.}
    \label{fig:V_M}
\end{figure}
 
As described in Section\,\ref{mathematics}, the viscosity function $\mu(c)$ is used to construct theoretical estimates for the size of mixing zone and corresponding velocities. Obviously, in addition to the shape of this function, a key role is played here by the viscosity ratio parameter $M$. To compare the simulation results with the estimates of interest, we used 4 different dependencies for $\mu$ on the polymer concentration -- concave, linear, quadratic, and exponential (see, Fig.\,\ref{fig:vis}). The comparison results of \eqref{eq:Koval} and \eqref{TFE-estimates} with the numerical solution are demonstrated in Fig.\,\ref{fig:V_M}. The bounds of the theoretical estimates for $V_1,V_2,V_3$ are highlighted in color. 
The simulations were performed for a 1/3\,PV polymer slug with a diffusion coefficient of $10^{-7}$\,m$^2$/s. The rest parameters are the same as in Tab.\,\ref{tab:param}. Note that, in general, there is a good coincidence between the COMSOL and MRST curves, but, as above, there is a slight difference in movement of the rear end of the slug.

\begin{figure}[t]
\centering
\includegraphics[width=0.49\textwidth]{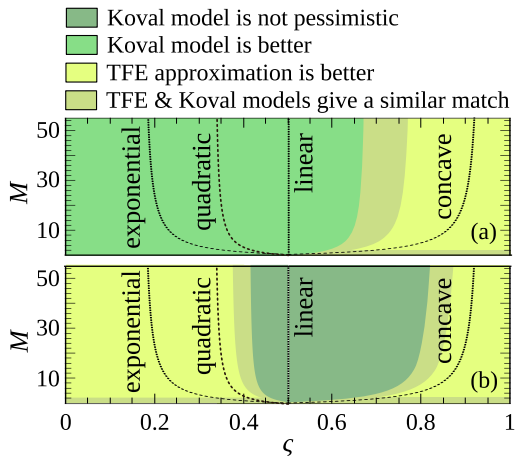}
\caption{The schematic representation of theoretical estimates that most closely matches the results of numerical modelling depending on the concavity parameter and the viscosity ratio for (a) front and (b) tail velocities of the mixing zone. The intermediate color shows areas where both models give identical estimates.}
\label{fig:fin}  
\end{figure}

For a generalization of the patterns reported in\,Fig.\,\ref{fig:V_M}, we introduce the parameter $\varsigma$ defining the concavity of the viscosity function as a ratio of the area under the curve $\mu(c)$ to the whole area of the containing square. In other words, we can write $\varsigma=\int_0^{c^\mathrm{max}}(\mu-\mu_1)\mathrm{d}c/[\mu_1c^\mathrm{max}(M-1)]$.
By taking into consideration that $\mu_1=200c^\mathrm{max}$, we get that $\varsigma=1/2$ for the linear viscosity function. For the rest of the analyzed viscosity functions in Tab.\ref{tab:param} the parameter $\varsigma$ occurs to be dependent on the viscosity ratio factor $M$. Figure\,\ref{fig:fin} depicts a summary diagram reflecting the most accurate method with reference to numerical results for a variety of viscosity ratio factors $M$ and concavity parameters $\varsigma$. The estimates for the front and tail velocities of the mixing zone are analyzed separately. The intermediate color shows areas where both models give identical output. Dark green color fills the area where Koval model is closer to simulations but is not pessimistic. 

One may conclude on the basis of the performed comparison that Koval estimates for the velocity of the front end of the mixing zone $V_2$ are more accurate than the TFE ones. In turn, for the rear end of the polymer slug $V_3$, the accuracy of the TFE and Koval approaches are comparable. 
The only exception from this observation is the concave viscosity. It should be emphasized that this shape is not physically justified and is presented only for mathematical validation of the models. From this, it can be summarised that the TFE estimates are more pessimistic for realistic viscosity functions in comparison with the Koval approach.

\subsubsection{When TFE approximation is exact?}
\label{sec:empmodels}
Analyzing Fig.\,\ref{fig:V_M}, we see that TFE estimates for the mixing zone velocities provide a reasonable fit to the numerical results in the following cases: concave viscosity (for the outermost finger), exponential viscosity (for the tailing end).
Let us give a theoretical reasoning for this effect. To derive the TFE approximation, we use the following inequalities

\begin{equation}
    m(1)\leqslant \overline{m}(x,t)\leqslant m(0).
    \label{estimates-velocity}
\end{equation}
Near the front end of the water/polymer interface equality $m(1)\approx \overline{m}(x,t)$ is fulfilled and near the rear end of the slug it is true that $m(0)\approx\overline{m}(x,t)$. Nevertheless, the estimates \eqref{estimates-velocity} are quite rough in the case of general viscosity function far from the boundaries of the mixing zone.
The more concave viscosity function we consider, the greater the region where $m(1)\approx \overline{m}(x,t)$ is, and the better TFE estimate for the outermost finger is. Analogously, the more convex viscosity function we consider, the greater the region  where $m(0)\approx \overline{m}(x,t)$ is, and the better TFE estimate for the tailing end is. 

\section{Conclusions}
\label{sec:summary}
We have numerically investigated the evolution of the mixing zone in one-phase miscible displacement in linear geometry corresponding to polymer flooding between two horizontal wells. The simulations were performed using two different solvers: COMSOL (finite elements) and MRST (finite volumes) and are in good agreement with each other. The water/polymer interface demonstrates a linear growth rate that corresponds to the constancy of movement velocities for mixing zone boundaries till the influence of the slug front prevails. At the same time, the change of the slug size does not affect results. Such a finding makes it possible to compare numerical results with that predicted by analytical methods. The theoretical estimates on velocity of the tip leading finger in the transverse flow equilibrium regime predicted by Otto \& Menon \cite{otto2006}, Yortsos \&Salin \cite{yortsos2006} are in agreement with the simulations. However, as we clearly demonstrated, this regime is too pessimistic in most cases. On the other hand, we found out that the growth rate of the mixing zone depends not only on the Peclet number and ratio of viscosities, but also on the form of the viscosity curve. This feature is captured by the TFE approximation, but not by the conventional Koval model. Crucially, the summary diagram may contain small inaccuracies because of the evaluation procedure for the determination boundary velocities. Hence, further research on the topic is encouraged to eliminate this drawback.

\section*{Acknowledgements}
\noindent
The authors acknowledge the support to this work by President Grant 075-15-2019-204, by "Native towns", a social investment program of PJSC "Gazprom Neft" and Ministry of Science and Education of Russian Federation, grant 075–15–2019–1619.

\bibliography{mybibfile}

\begin{thebibliography}{10}
\expandafter\ifx\csname url\endcsname\relax
  \def\url#1{\texttt{#1}}\fi
\expandafter\ifx\csname urlprefix\endcsname\relax\def\urlprefix{URL }\fi
\expandafter\ifx\csname href\endcsname\relax
  \def\href#1#2{#2} \def\path#1{#1}\fi

\bibitem{bakharev2020}
F.~Bakharev, L.~Campoli, A.~Enin, S.~Matveenko, Y.~Petrova, S.~Tikhomirov,
  A.~Yakovlev, Numerical investigation of viscous fingering phenomenon for raw
  field data, Transport in Porous Media (2020) 1--22.

\bibitem{reviewmiscible2017}
G.~Scovazzi, M.~F. Wheeler, A.~Mikeli{\'c}, S.~Lee, Analytical and variational
  numerical methods for unstable miscible displacement flows in porous media,
  Journal of Computational Physics 335 (2017) 444--496.

\bibitem{homsy1987}
G.~M. Homsy, Viscous fingering in porous media, Annual review of fluid
  mechanics 19~(1) (1987) 271--311.

\bibitem{chen1998part1}
C.-Y. Chen, E.~Meiburg, Miscible porous media displacements in the quarter
  five-spot configuration. part 1. the homogeneous case, Journal of Fluid
  Mechanics 371 (1998) 233--268.

\bibitem{chen1998part2}
C.-Y. Chen, E.~Meiburg, Miscible porous media displacements in the quarter
  five-spot configuration. part 2. effect of heterogeneities, Journal of Fluid
  Mechanics 371 (1998) 269--299.

\bibitem{ChuokeMeursPoel5}
R.~Chuoke, P.~Van~Meurs, C.~van~der Poel, et~al., The instability of slow,
  immiscible, viscous liquid-liquid displacements in permeable media, Petroleum
  Trans., AIME 216 (1959) 188--194.

\bibitem{luo2018scaling}
H.~Luo, M.~Delshad, G.~A. Pope, K.~K. Mohanty, Scaling up the interplay of
  fingering and channeling for unstable water/polymer floods in viscous-oil
  reservoirs, Journal of Petroleum Science and Engineering 165 (2018) 332--346.

\bibitem{luo2017interactions}
H.~Luo, M.~Delshad, G.~A. Pope, K.~K. Mohanty, et~al., Interactions between
  viscous fingering and channeling for unstable water/polymer floods in heavy
  oil reservoirs, in: SPE Reservoir Simulation Conference, Society of Petroleum
  Engineers, 2017, p.~26.

\bibitem{Koval15}
E.~Koval, et~al., A method for predicting the performance of unstable miscible
  displacement in heterogeneous media, Society of Petroleum Engineers Journal
  3~(02) (1963) 145--154.

\bibitem{otto2005}
G.~Menon, F.~Otto, Dynamic scaling in miscible viscous fingering,
  Communications in mathematical physics 257~(2) (2005) 303--317.

\bibitem{yortsos2006}
Y.~Yortsos, D.~Salin, On the selection principle for viscous fingering in
  porous media, Journal of Fluid Mechanics 557 (2006) 225.

\bibitem{ToddLongstaff16}
M.~Todd, W.~Longstaff, et~al., The development, testing, and application of a
  numerical simulator for predicting miscible flood performance, Journal of
  Petroleum Technology 24~(07) (1972) 874--882.

\bibitem{Fayers17}
F.~J. Fayers, et~al., An approximate model with physically interpretable
  parameters for representing miscible viscous fingering, SPE reservoir
  engineering 3~(02) (1988) 551--558.

\bibitem{TardyPearson18}
P.~Tardy, J.~Pearson, A 1d-averaged model for stable and unstable miscible
  flows in porous media with varying peclet numbers and aspect ratios,
  Transport in porous media 62~(2) (2006) 205--232.

\bibitem{wooding1969}
R.~A. Wooding, Growth of fingers at an unstable diffusing interface in a porous
  medium or hele-shaw cell, Journal of Fluid Mechanics 39~(3) (1969) 477--495.

\bibitem{otto2006}
G.~Menon, F.~Otto, Fast communication: Diffusive slowdown in miscible viscous
  fingering, Communications in Mathematical Sciences 4~(1) (2006) 267--273.

\bibitem{koval1963}
E.~Koval, et~al., A method for predicting the performance of unstable miscible
  displacement in heterogeneous media, Society of Petroleum Engineers Journal
  3~(02) (1963) 145--154.

\bibitem{booth2008phd}
R.~Booth, Miscible flow through porous media, PhD thesis, University of Oxford,
  2008.

\bibitem{booth2010}
R.~Booth, On the growth of the mixing zone in miscible viscous fingering,
  Journal of fluid mechanics 655 (2010) 527.

\bibitem{toddlongstaff1972}
M.~Todd, W.~Longstaff, et~al., The development, testing, and application of a
  numerical simulator for predicting miscible flood performance, Journal of
  Petroleum Technology 24~(07) (1972) 874--882.

\bibitem{Oswald97}
S.~Oswald, W.~Kinzelbach, A.~Greiner, G.~Brix, Observation of flow and
  transport processes in artificial porous media via magnetic resonance imaging
  in three dimensions, Geoderma 80~(3-4) (1997) 417--429.

\bibitem{Gupta74}
S.~P. Gupta, R.~A. Greenkorn, An experimental study of immiscible displacement
  with an unfavorable mobility ratio in porous media, Water Resources Research
  10~(2) (1974) 371--374.

\bibitem{KeyesMcInnesWoodwardGroppMyraPerniceWohlmuth2013}
D.~E. Keyes, L.~C. McInnes, C.~Woodward, W.~Gropp, E.~Myra, M.~Pernice,
  B.~Wohlmuth, Multiphysics simulations: Challenges and opportunities, The
  International Journal of High Performance Computing Applications 27~(1)
  (2013) 4--83.

\bibitem{Comsol38}
R.~W. Pryor, Multiphysics modeling using COMSOL{\textregistered}: a first
  principles approach, Jones \& Bartlett Publishers, 2009.

\bibitem{MRST}
K.-A. Lie, An introduction to reservoir simulation using MATLAB/GNU Octave:
  User guide for the MATLAB Reservoir Simulation Toolbox (MRST), Cambridge
  University Press, 2019.

\bibitem{bao2017}
K.~Bao, K.-A. Lie, O.~M{\o}yner, M.~Liu, Fully implicit simulation of polymer
  flooding with mrst, Computational Geosciences 21~(5-6) (2017) 1219--1244.

\bibitem{Persson05}
P.~Persson, G.~Strang, A simple mesh generator in matlab, SIAM Review 46~(2)
  (2004) 329--345.

\bibitem{Nijjer}
J.~S. Nijjer, D.~R. Hewitt, J.~A. Neufeld, The dynamics of miscible viscous
  fingering from onset to shutdown, Journal of Fluid Mechanics 837 (2018)
  520--545.

\bibitem{Orr07}
F.~M. Orr, et~al., Theory of gas injection processes, Vol.~5, Tie-Line
  Publications Copenhagen, 2007.

\end{thebibliography}

\end{document}